\documentclass[prb,twocolumn,superscriptaddress]{revtex4}

\usepackage{graphicx}
\usepackage{amsmath}
\usepackage{hyperref}

\begin{document}


\title{Manipulating Mn--Mg$_k$ cation complexes \\ to control the charge- and spin-state of Mn in GaN}

\author{Thibaut Devillers}
\email{thibaut.devillers@jku.at}

\affiliation{Institut f\"ur Halbleiter-und-Festk\"orperphysik, Johannes Kepler University, Altenbergerstr. 69, A-4040 Linz, Austria}

\author{Mauro Rovezzi}
\affiliation{European Synchrotron Radiation Facility, 6 rue Jules Horowitz, F-38043 Grenoble, France}

\author{Nevill Gonzalez Szwacki}
\affiliation{Institute of Theoretical Physics, Faculty of Physics, University of Warsaw, ul. Ho\.za 69, PL-00-681 Warszawa, Poland}

\author{Sylwia Dobkowska}
\affiliation{Institute of Physics, Polish Academy of Sciences, al. Lotnik\'{o}w 32/46, PL-02-668 Warszawa, Poland}

\author{Wiktor Stefanowicz}
\affiliation{Institute of Physics, Polish Academy of Sciences, al. Lotnik\'{o}w 32/46, PL-02-668 Warszawa, Poland}

\author{Dariusz Sztenkiel }
\affiliation{Institute of Physics, Polish Academy of Sciences, al. Lotnik\'{o}w 32/46, PL-02-668 Warszawa, Poland}

\author{Andreas Grois}
\affiliation{Institut f\"ur Halbleiter-und-Festk\"orperphysik, Johannes Kepler University, Altenbergerstr. 69, A-4040 Linz, Austria}

\author{Jan Suffczy\'{n}ski}
\affiliation{Institute of Experimental Physics, Faculty of Physics, University of Warsaw, ul. Ho\.za 69, PL-00-681 Warszawa, Poland}

\author{Andrea Navarro-Quezada}
\affiliation{Institut f\"ur Halbleiter-und-Festk\"orperphysik, Johannes Kepler University, Altenbergerstr. 69, A-4040 Linz, Austria}

\author{Bogdan Faina}
\affiliation{Institut f\"ur Halbleiter-und-Festk\"orperphysik, Johannes Kepler University, Altenbergerstr. 69, A-4040 Linz, Austria}

\author{Tian Li}
\affiliation{Institut f\"ur Halbleiter-und-Festk\"orperphysik, Johannes Kepler University, Altenbergerstr. 69, A-4040 Linz, Austria}

\author{Pieter Glatzel}
\affiliation{European Synchrotron Radiation Facility, 6 rue Jules Horowitz, F-38043 Grenoble, France}

\author{Francesco d'Acapito}
\affiliation{Consiglio Nazionale delle Ricerche, IOM-OGG, c/o ESRF GILDA CRG, BP 220, F-38043 Grenoble, France}

\author{Rafa\l{} Jakie\l{}a}
\affiliation{Institute of Physics, Polish Academy of Sciences, al. Lotnik\'{o}w 32/46, PL-02-668 Warszawa, Poland}

\author{Maciej Sawicki}
\affiliation{Institute of Physics, Polish Academy of Sciences, al. Lotnik\'{o}w 32/46, PL-02-668 Warszawa, Poland}

\author{Jacek~A.~Majewski}
\affiliation{Institute of Theoretical Physics, Faculty of Physics, University of Warsaw, ul. Ho\.za 69, PL-00-681 Warszawa, Poland}

\author{Tomasz Dietl}
\affiliation{Institute of Physics, Polish Academy of Sciences, al. Lotnik\'{o}w 32/46, PL-02-668 Warszawa, Poland}
\affiliation{Institute of Theoretical Physics, Faculty of Physics, University of Warsaw, ul. Ho\.za 69, PL-00-681 Warszawa, Poland}

\author{Alberta Bonanni}
\email{alberta.bonanni@jku.at}
\affiliation{Institut f\"ur Halbleiter-und-Festk\"orperphysik, Johannes Kepler University, Altenbergerstr. 69, A-4040 Linz, Austria}



\begin{abstract}

Owing to the variety of possible charge and spin states and to the different ways of coupling to the environment, paramagnetic centres in wide band-gap semiconductors and insulators exhibit a strikingly rich spectrum of properties and functionalities, exploited in commercial light emitters and proposed for applications in quantum information. Here we demonstrate, by combining synchrotron techniques with magnetic, optical and \emph{ab initio} studies, that the codoping of GaN:Mn with Mg allows to control the Mn$^{n+}$ charge and spin state in the range $3$\,$\le$\,$n$\,$\le$\,$5$ and $2$\,$\ge$\,$S$\,$\ge$\,$1$. According to our results, this outstanding degree of tunability arises from the formation of hitherto concealed cation complexes Mn-Mg$_k$, where the number of ligands $k$ is pre-defined by fabrication conditions. The properties of these complexes allow to extend towards the infrared the already remarkable optical capabilities of nitrides, open to solotronics functionalities, and generally represent a fresh perspective for magnetic semiconductors.

\end{abstract}

\maketitle


Group III nitrides currently dominate the field of visible and ultraviolet photonics, due to the flexibility of gap engineering, to the availability of both $n$- and $p$-type material, to their high thermal stability, and large heat conductivity (ref.~\onlinecite{Morkoc:2009_book} and references therein).
Moreover, recent studies of magnetically doped semiconductors\cite{Sato:2010_RMP,Bonanni:2010_CSR,Jamet:2006_NM,Yokoyama:2005_JAP} and oxides\cite{Rode:2008_APL,Kim:2005_PRB} have pointed out how crucial is the spatial distribution of dopants in the host matrix. In particular, the aggregation of magnetic impurities resulting at the nanoscale in the formation of chemical and/or crystallographic inhomogeneities has a dramatic influence on the magnetic, electronic and optical properties of the systems. The possibility to reach a control over the formation, structure, arrangement, and effect of inhomogeneities on the chemical and physical behavior of the host material, can be envisaged to have outstanding consequences in the design and functionality of the next generation of devices. The manipulation of the charge-state of paramagnetic centres in wide band gap semiconductors by modifying the local electric field with a gate voltage\cite{Grotz:2012_NC} or chemically by changing the termination of the surface\cite{Hauf:2011_PRB}, has already been reported.

Here, we demonstrate the efficiency of an alternative approach based on the controlled formation of complexes, involving one magnetic impurity and one or more electrically active dopant. Peculiar arrangements in the electronic structure of simple paramagnetic complexes have previously been theoretically investigated\cite{Raebiger:2010_PRB}.
With synchrotron radiation methods supported by \emph{ab initio }computations, we show that codoping of GaN with Mn and Mg results in the formation of cation complexes Mn--Mg$_k$. Depending on the number of ligands $k$ pre-defined by fabrication conditions, a strong electron-phonon interaction and consequently an efficient and broadband infrared (IR) photoluminescence (PL) are promoted. Furthermore, we find that the Mn--Mg$_k$ cation complexes allow to control the charge and spin states of the transition metal (TM) ions, with prospective implications in the design of the coupling between localized spins in magnetic semiconductors and in the optimisation of centres for solotronic applications\cite{Koenraad:2011_NM}.


\section{Results}

The samples studied are single-crystal GaN layers, codoped with Mn and Mg, each with concentrations lower than 1\%. Their fabrication,architecture, and preliminary characterisation are summarised in the \emph{Methods}. 
Aware that the incorporation of interstitial hydrogen forming complexes with Mg is a recurrent challenge associated with Mg-doping of GaN, we have performed a careful chemical analysis, detailed in the Supplementary Figure~S1, showing that in our case we can reasonably neglect the effect of interstitial hydrogen.


\subsection{Nature of Mn--Mg complexes}

\begin{figure*}[ht]
	\centering
  \includegraphics[width=0.95\textwidth]{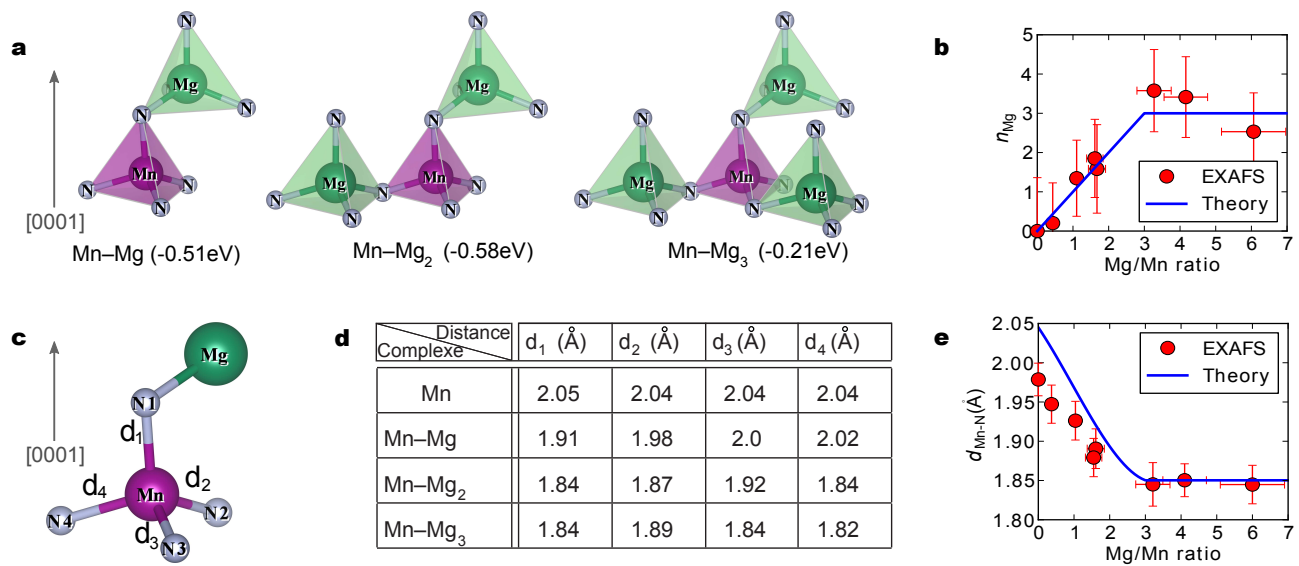}
	\caption{   \label{fig:complexes}\textbf{Mn--Mg$_k$ complexes predicted by theory and experimental demonstration by EXAFS.} \textbf{a}, most stable Mn--Mg$_k$ complexes ($k$\,=\,1,2 and 3) and their pairing energies -- computed by DFT within the GGA+U approximation -- relatively to the previous Mn-Mg$_{k-1}$ complex. A gray arrow indicates the [0001] direction, \emph{i.e.} the GaN \emph{c}-axis; \textbf{b}, number of Mg atoms seen by Mn in the first cation coordination sphere ($n_\mathrm{Mg}$), extracted from EXAFS	measurements and DFT predictions; \textbf{c}, schematic representation of a Mn--Mg complex;  \textbf{d}, detailed values of the bond lengths for different complexes, calculated by DFT; \textbf{e}, average Mn-N bond length ($d_\mathrm{Mn-N}$) as a function of the Mg/Mn ratio, from EXAFS analysis and DFT calculations.}

\end{figure*}

Having established the single crystallinity and chemical homogeneity of the samples $via$ a spectrum of both local and averaging characterisation techniques, we discuss the lattice positions of Mn and Mg impurities in GaN, as determined from extended x-ray absorption fine structure (EXAFS) measurements at the Mn K-edge and by {\em ab initio} computations. We find that within a confidence of 90\% the Mn ions occupy exclusively Ga-substitutional positions. This result is consistent with the {\em ab initio} computations reported in the Supplementary Table~S1 showing that the energy required to place a Mn or Mg ion in either octahedral or tetrahedral interstitial positions of GaN is more than 4\,eV higher than the one needed to incorporate it in a Ga-substitutional site.

According to our previous EXAFS and electron energy loss spectroscopy (EELS) results, Mn is randomly and homogeneously distributed in GaN:Mn, at least up to a Mn concentration of 3\%~(refs. \onlinecite{Stefanowicz:2010_PRBa,Bonanni:2011_PRB}). 
While in the conventional treatment of dilute magnetic semiconductors, the spatial distributions of co-dopants and TM ions are assumed to be uncorrelated, a quantitative analysis of our EXAFS data points to a substantial \emph{correlation} between the positions occupied by Mn and Mg in the host lattice.  Simulations of EXAFS spectra for a large variety of relaxed defects in GaN were performed and reported in the Supplementary Information, and indicate that the combination of substitutional Mn and Mg is the most likely to account for the experimental EXAFS data. The experimental EXAFS spectra are then fitted, according to the procedure described in the Supplementary Information. As shown in Fig.~\ref{fig:complexes}b, the number of substitutional Mg atoms ($n_\mathrm{Mg}$) in the first cation coordination sphere of Mn increases linearly with the ratio between the Mg and Mn concentrations, $y = x_{\mathrm{Mg}}/x_{\mathrm{Mn}}$, up to $y$\,$=$\,$3$ and then saturates at higher $y$ values. Simultaneously, the average distance beween Mn and the nearest neighbour N atoms $d_\mathrm{Mn-N}$ diminishes in range up to $y=3$ and then levels off, as seen in Fig.\,\ref{fig:complexes}e. The experimental EXAFS spectra from which $d_\mathrm{Mn-N}$ and $n_\mathrm{Mg}$ are extracted are plotted in Supplementary Figure~S4 for different values of $y$. It is possible to confirm the correlation between the Mn and Mg positions by comparing the spectra for $y>0$ (correlated) and $y=0$, the latter being strictly equivalent to the \emph{not} correlated case where Mn does not interact with any Mg atom.

These results can be explained by our {\em ab initio} computations. In particular, the estimated values of the pairing energies $E_{\text{p}}$ shown in Fig.\,\ref{fig:complexes}a for up to three Mg per Mn are all negative, demonstrating the tendency of Mg to form the Mn-Mg$_{k}$ complexes sketched in Fig.\,\ref{fig:complexes}a and in Supplementary Figure\,S8. Taking into account the statistical distribution of $k$, implying that some of these complexes can coexist at a given $y$, we have obtained with no adjustable parameters a remarkable agreement between the experimental and computed trends, describing how the number of bound Mg atoms (Fig.\,\ref{fig:complexes}b) and the shortening of the bond length (Fig.\,\ref{fig:complexes}c,d,e) depend on $y$. This agreement with the theory based on the statistical distribution of the complexes populations detailed in the \emph{Methods} implies a comparable stability of the complexes with different $k$, up to $k=3$. According to the DFT calculations and as reported in Fig. 1a, a similar binding energy (0.5\,eV) is indeed computed for Mn-Mg and Mn-Mg$_2$ but a lower value (0.2\,eV) is expected for Mn-Mg$_{3}$. If these DFT predictions are quantitatively correct, we may await a concentration of Mn-Mg$_3$ complexes somewhat lower than the one resulting from $P_3(y)$ for $y<1$.
According to the \emph{ab initio} results in Fig.~\ref{fig:complexes}d, the bond shortening is particularly significant for the Mn-N pairs nearest to Mg. Intuitively, this effect results from a sizable charge transfer at the Mn-Mg$_k$ complexes, that leads to the onset of Coulomb attraction between the ionized Mg acceptors and Mn donor.


\subsection{Control of charge and spin state}

\begin{figure*}[htb]
  \centering
  \includegraphics[width=0.87\textwidth]{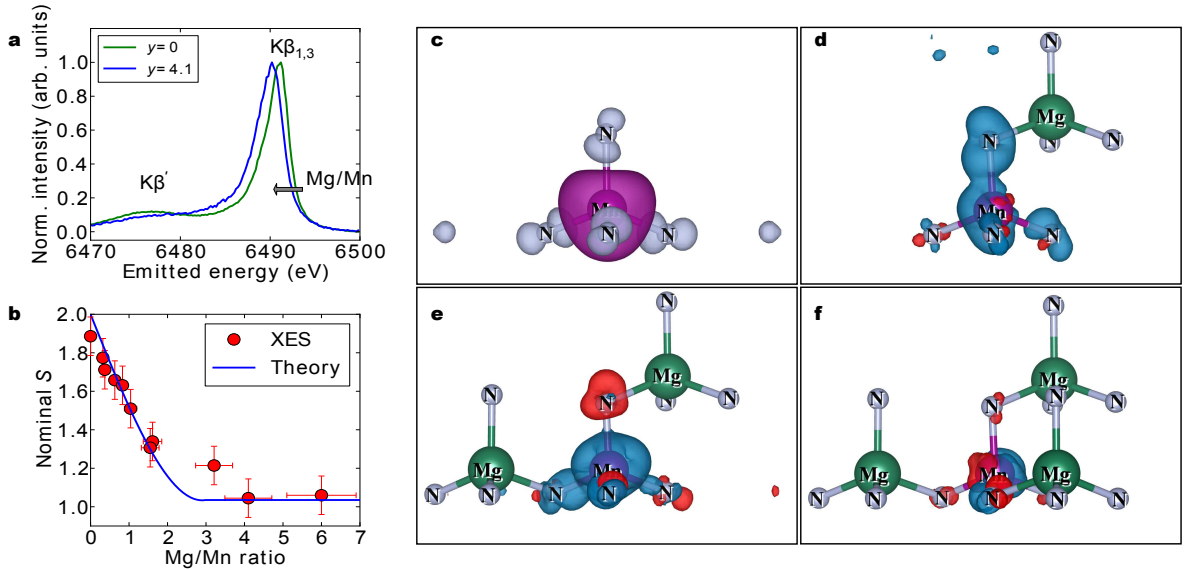}
  \caption{\label{fig:spin} \textbf{Evolution of the Mn spin state}. \textbf{a}, non-resonant XES for GaN:Mn samples with and without Mg;	\textbf{b}, evolution of the nominal spin value with the ratio between Mg and Mn concentration, extracted from the analysis of XES spectra  and calculated \emph{via} DFT; \textbf{c}, computed spin polarization density [$\Delta \rho =\rho(\uparrow)-\rho(\downarrow)$] for Mn$_\mathrm{Ga}$ in wurtzite (wz)-GaN. The positive and negative spin polarizations are represented by violet and grey colours, respectively; \textbf{d--f}, difference $\rho_k-\rho_{k-1}$, between the spin polarizations of the Mn--Mg$_k$ complexes, with $k=1$, 2 and 3. The red and blue colours represent positive and negative values, respectively. The blue colour in {\bf d} indicates the enhanced delocalization of spin polarization in Mn-Mg, whereas the red colour in {\bf e} and {\bf f} points to a gradual shift of the spin density to Mn in Mn-Mg$_2$ and then in Mn-Mg$_3$.  In all plots the contour value is set to 0.005 corresponding to about $2\times10^{-5} \mu_\mathrm{B}\,$\AA$^{-3}$.}

\end{figure*}

Having discussed the atomistic structure of the Mn-Mg$_{k}$ complexes, we determine their charge and spin states. According to our {\em ab initio} computations, if Mn-Mg$_k$ complexes were not formed, only a single hole would be trapped by Mn upon Mg codoping, resulting in the Mn spin state $S$\,=\,3/2.  In that case, if $y$\,$>$\,$1$,  extra holes would be directed to the valence band, increasing the \emph{p}-type conductivity. However, layers containing a small concentration of Mg ($y<3$) are insulating and the onset of conductivity takes place only for $y\ge 3$, with a decrease in the resistance of the layers of more than three orders of magnitude, indicating that at low $y$, Mg are bound to Mn in complexes, and do not exist as isolated acceptors, as free Mg acceptors are only present for $y \ge 3$. Considering the Mn-Mg$_k$ complexes, our computations predict the presence of Mn with spin state down to $S$\,=\,1 for $k \ge 2$. Altogether, our theoretical studies suggest the possibility to control, with Mg codoping, the Mn$^{n+}$ charge and spin state over the range $3  \le n \le 5$ and $2 \ge S \ge 1$, respectively.

We give now experimental verification of these theoretical expectations by tracing the evolution of $S$ upon Mg codoping.
We exploit the K$\beta$ x-ray emission spectroscopy (XES) technique which, by probing the 3$p$ $\to$ 1$s$ transitions, is sensitive to the magnitude of the exchange interaction between the Mn 3$p$ core-hole and the net magnetic moment in the Mn 3$d$ valence shell \cite{Tsutsumi:1976_PRB}.
We note that compared to x-ray absorption spectroscopy (XAS), where a charge state dependent shift of the absorption edge is also visible, the XES data are more linearly correlated with the spin state, and depend less on the atomic configuration\cite{Pizarro:2004_PCCP}.
The spectra are plotted in Fig.~\ref{fig:spin}a from which we extract, by comparison with reference oxides, as described in the Supplementary Information, the values of $S$ reported in Fig.~\ref{fig:spin}b. This analysis demonstrates that upon Mg codoping the Mn spin state first decreases linearly from 2 to 1, and then saturates for $y \gtrsim 3$. As shown in Fig.~\ref{fig:spin}b, these results are in accord with the magnitudes of $S$ determined from the {\em ab initio} computations for a unit cell containing one Mn.


\subsection{Magnetism of complexes}

Having established the TM charge and spin states, one can examine the properties of the centres and their coupling to the environment in terms of the time-honoured crystal-field theory\cite{Henderson:2006_B}, providing the expected structure and symmetries of the relevant energy levels at a given oxidation state\cite{Jansen:2008_ACh}. In order to gain more understanding at the microscopic level\cite{Raebiger:2008_N}, it is instructive to consider the actual spin distribution around the Mn ions and its evolution with Mg codoping. As shown in Fig.~\ref{fig:spin}b and in the Supplementary Table~S2, our \emph{ab initio} computations confirm that in the case of GaN:Mn the total magnetic moment $m$\,=\,4\,$\mu_{\text{B}}$ is built of $m_{\text{c}}$\,$\approx$\,4.5\,$\mu_{\text{B}}$ on the $d$ shell of the Mn ion and  $m_{\text{a}}$\,$\approx$\,-0.5\,$\mu_{\text{B}}$ on the $p$ orbitals of the N ligands. This means, in agreement with the notion of strong coupling limit for GaN:Mn (ref.~\onlinecite{Dietl:2008_PRB}), that the wave function of the hole provided by the Mn is equally spread between Mn and the neighbouring N ions.
As reported in Fig.~\ref{fig:spin}d, the delocalization of holes is enhanced for $k$\,=\,1, suggesting that Mn-Mg complexes can mediate spin-dependent interactions between Mn spins.  On the other hand, as evidenced in Fig.~\ref{fig:spin}e,f, for $k$\,=\,2 and 3 the delocalization is quenched, and this result is important for understanding the optical data discussed later.

The changes of the Mn spin state and Mn-N bonding with Mg codoping, evaluated by EXAFS, XES, and {\em ab initio} studies affect the magnetic properties of the system. Quite generally, the values of the magnetization $M(H)$ and its anisotropy are determined by the relevant spin Hamiltonian, whose form, appropriate for a given $S$, is determined by crystal symmetry, including local strains, whereas the magnitudes of the spin Hamiltonian parameters provide information on the coupling to the environment (including $p-d$ hybridization) and on the strength of the spin-orbit interaction. In GaN:Mn, a sizable difference in the magnetization curves $M(H)$ is observed for two orientations of the magnetic field $H$ with respect to the wurtzite $c$-axis, $H \perp c$ and $H\parallel c$ (refs.~\onlinecite{Stefanowicz:2010_PRBa,Gosk:2005_PRB}). A quantitative analysis of such data allowed to obtain the values of the spin Hamiltonian parameters for this case, with $S =2$ and wurtzite symmetry \cite{Stefanowicz:2010_PRBa,Gosk:2005_PRB}. In the case of GaN:(Mn,Mg), as seen in Fig. 1c-d, the distortions of the nitrogen tetrahedron and, in particular, the shortening of the Mn-N bond induced by the presence of Mg, is much more pronounced than the elongation of the Mn-N bond parallel to the $c$-axis. Accordingly, the local anisotropy will now be determined by the position of one or more Mg in the first cation coordination sphere. Since there is no preferential orientation for the occupation of any of the 12 nearest neighbour positions, the local anisotropy and, thus, the orientation of the easy axis will be randomly distributed over the Mn-Mg$_k$ complexes. Hence, the presence of Mg leads to the disappearance of anisotropy with respect to the $c$-axis. Indeed, as shown in Fig. 3, we observe that the magnitude $A$ of the magnetic anisotropy decays to zero with increasing $y$. In agreement with the model, $A(y)$  follows the probability $P_0(y)$ that no Mg is bound by Mn at a given $y$, $A(y) = A(0)(1-y/3)^3$. At the same time, the absolute values of $M(H)$ are determined by unknown parameters of the spin Hamiltonians, appropriate for particular values of $S$, local strains, and strength of the spin-orbit interaction.

\begin{figure}[htb]
	\centering
  \includegraphics[width=0.45\textwidth]{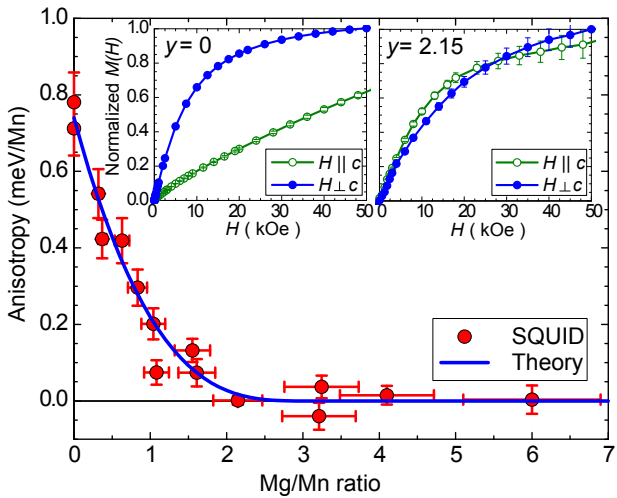}
  \caption{  \label{fig:SQUID}\textbf{Magnetism of Mn--Mg$_k$ complexes.} Magnetic anisotropy energy density as a function of $y$ measured by SQUID magnetometry, and calculated assuming that it originates from Mn without Mg in the first cation coordination sphere. Inset: normalized magnetization curves $M$--$H$ of GaN:Mn containing 0.4\% of Mn without and with Mg codoping (left and right panel, respectively) measured at 1.85~K.}

\end{figure}


\begin{figure*}[htb]
	\centering
  \includegraphics[width=0.98\textwidth]{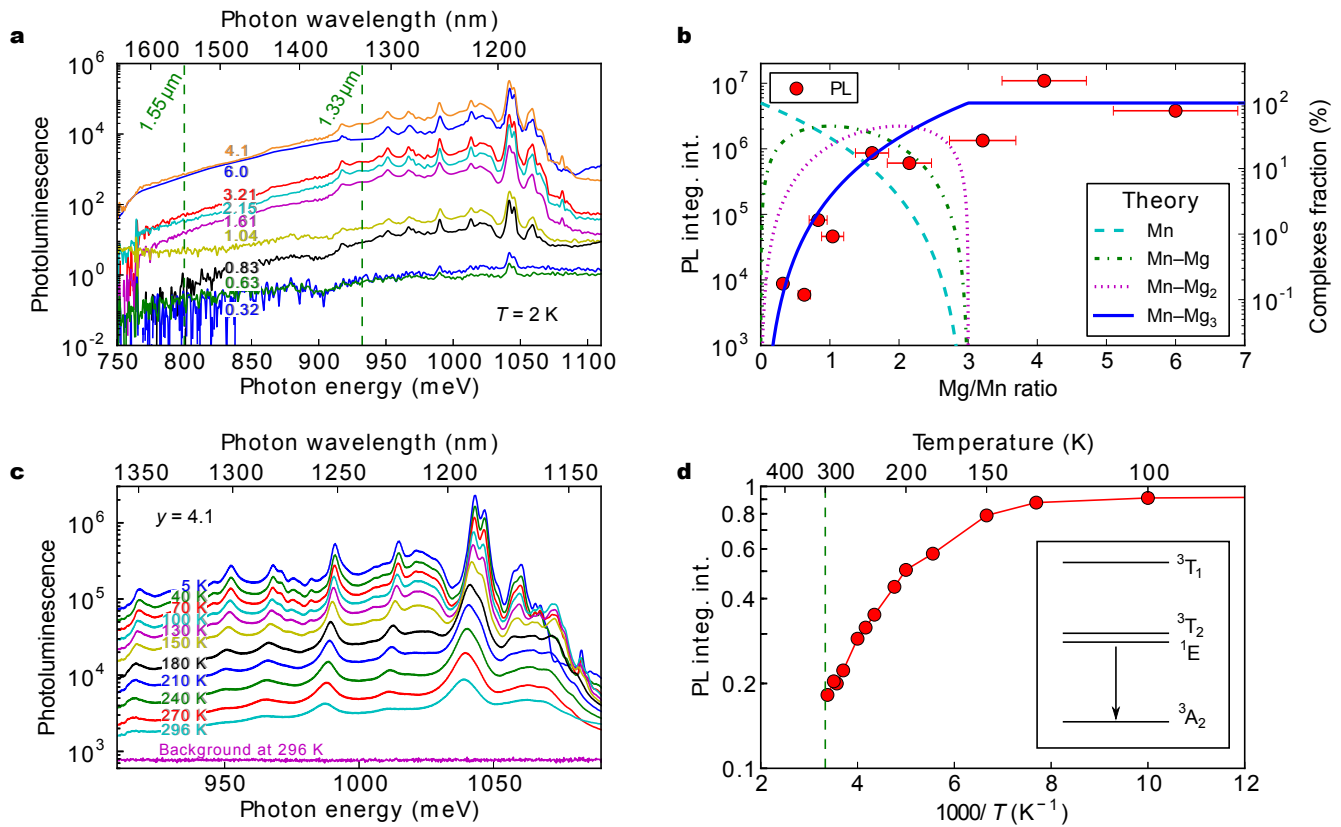}
  \caption{  \label{fig:PL}\textbf{Infrared photoluminescence of GaN:(Mn,Mg) samples.} \textbf{a}, evolution of the PL spectra (excited with a  442 nm (2.8 eV) laser)  as a function of the $y$ ratio, measured at 2~K (a multiplying factor of 1.4 has been applied between consecutive spectra for clarity); \textbf{b}, integrated PL intensity normalized by the Mn concentration and sample thickness as a function of the $y$ ratio (red circles, left scale) and evolution of the fraction of different complexes, calculated as a function of $y$ (lines, right scale); \textbf{c}, evolution of the PL spectra with temperature, for a sample with $y=4.1$; the background at 296 K is also included. A factor of 1.4 has been applied between consecutive spectra for clarity. \textbf{d}, evolution of the integrated PL intensity as a function of the inverse temperature, normalized to the value at $T$\,=\,5\,K; the line is a guide for the eye. Inset: levels relevant to PL as discussed in the Supplementary Information. }

\end{figure*}

\subsection{Effect of complexes on the luminescence of Mn}

Particularly remarkable is the influence of the Mn--Mg$_k$ complexes on the light emission of GaN. At present, TM-doping of oxides and II-VI chalcogenides allows the fabrication of optically pumped tunable lasers\cite{Sorokin:2005_IEEE,Mirov:2010_LPR}.

The PL spectra reported in Fig.~\ref{fig:PL}a for our samples, extends for high $y$-ratios over a broad infrared band, covering two of the telecommunication windows, namely 1.33\,$\mu$m and 1.55\,$\mu$m. In addition, the multisite structure suggests that only one transition is involved, but different environments as well as phonon replicas are responsible for the peaks multiplicity. The overall shape is indicative of a Huang-Rhys factor $S_{\text{HR}}$\,$>$\,2. According to Fig.\,\ref{fig:PL}b, the PL intensity increases with the concentration of the Mn-Mg$_3$ complexes  $P_3(y) = (y/3)^3$, demonstrating that Mn$^{5+}$, in presumably two slightly different environments, accounts for the photoluminescence in GaN:(Mn,Mg).

Importantly, as one sees in Fig~\ref{fig:PL}c and d, the spectrally broad infrared emission persists up to room temperature, and is very attractive for ultrashort pulse generation as well as for wide infrared tunability.

In order to understand the photoluminescence related to Mn--Mg$_k$ cation complexes, it is necessary to consider first the reason of the poor luminescence of Mn$^{3+}$ in GaN.
In contrast to Cr$^{2+}$ in Al$_2$O$_3$ and ZnSe, Mn$^{3+}$ in GaN does not show the strong and application-relevant red or infrared emission associated with optical transitions between the $^5$E and $^5$T$_2$ crystal field multiplets of the 3$d^4$ shell. This surprising result is however in agreement with previous extensive studies of GaN:Mn (refs~\onlinecite{Graf:2002_APL,Marcet:2006_PRB,Zenneck:2007_JAP,Malguth:2008_MRS}), in which the intra-centre photoluminescence of Mn$^{3+}$ was found to be hardly detectable. This puzzling result can be traced back to an abnormally small value of the Huang-Rhys factor $S_{\text{HR}} \lesssim 1$ implied by optical absorption\cite{Wolos:2004_PRBb,Marcet:2006_PRB}, this observation being interpreted\cite{Dietl:2008_PRB} in terms of a strong $p$-$d$ coupling in GaN:Mn, leading to a significant admixture of anion $t_2$ orbitals with the wave functions of the $^5$E and $^5$T$_2$ states. Such delocalization of the centre wave function, clearly visible in Fig.~\ref{fig:spin}c, reduces the electron-phonon coupling. As a result, the oscillator strength is shifted to the zero-phonon line at the expense of the phonon-assisted transitions that cease to constitute the channel for PL excitation. 

On the other hand, highly efficient and Stokes shifted PL, similar to the one presented in Fig.~\ref{fig:PL}, was found for GaN:(Mn,Mg) and assigned to intra Mn$^{4+}$ transitions\cite{Malguth:2008_MRS,Korotkov:2001_PBb,Han:2005_APL}. 
However, we have demonstrated here that following the evolution of the PL as a function of the different populations of complexes (Fig.~\ref{fig:PL}b), the Mn$^{5+}$ present in the complexes are likely to be at the origin of the IR PL signal.
This assignment is consistent with the shape of the PL spectrum in a 10~T magnetic field -- given in the Supplementary Figure~S7 --  which can be described by a splitting of the ground state into three components. This splitting is expected for $S$\,=\,1, as observed previously\cite{Thurian:1997_APL} for V$^{3+}$ in AlN.

In order to explain the origin of the PL activation upon codoping with Mg evidenced in Fig.~\ref{fig:PL}a, we refer to Fig.~\ref{fig:spin}d-f where the variations in the local spin densities brought about by the binding of an increasing number of Mg ions are shown. According to these data, the complexing with one Mg ion enhances the delocalization of the spin density over neighbouring N anions. However, with the binding of two and then three Mg ions, the
delocalization of the spin density decreases. Accordingly, a strong electron-phonon coupling and, thus, a large magnitude of $S_{\text{HR}}$ is restored, particularly for Mn--Mg$_3$ complexes, where two $d$ electrons reside in the $e$ orbitals of the $^3$A$_2$ ground state.

This IR broadband PL promoted by  Mn--Mg$_k$ cation complexes is of high relevance in \emph{e.g.} laser and telecommunication technologies. Actually, in comparison with Al$_2$O$_3$ and ZnSe, GaN has respectively seven and twelve times better thermal conductivity, lessening thermal effects even for high laser powers and intensities.


\section{Discussion}

We conclude that the data presented here indicate a new way to manipulate the charge and spin state of single paramagnetic centres by complexing a magnetic impurity with electrical dopants. The demonstration of these new degrees of freedom opens wide prospects illustrated here by the infrared emission of GaN:(Mn,Mg). Another line of research is to explore the potential of these Mn--Mg$_k$ cation complexes for mediating the coupling between localized spins in magnetic semiconductors. Furthermore, these centres may serve for storing and manipulating information in a single qubit or for single photon generation\cite{Koenraad:2011_NM}. Interestingly, unlike the case of CdTe:Mn (ref.~\onlinecite{Besombes:2004_PRL}) or InAs:Mn (ref.~\onlinecite{Krebs:2009_PRB}), it is not necessary to place GaN:(Mn,Mg) in a quantum dot, as Mn in GaN can bind the exciton\cite{Suffczynski:2011_PRB} needed to read or write information. The possibility to change energy level splitting, excitation channel and excited state lifetime by manipulating the Mn charge and spin state through Mn--Mg$_k$ cation complexes offers a not yet explored spectrum of opportunities for further investigations.


\section{Methods}

\textbf{ Growth and preliminary characterisation}: The samples consist of single crystal wurtzite (wz) GaN codoped with Mn and Mg grown by metalorganic vapor phase epitaxy (MOVPE) on a 1~$\mu$m GaN buffer layer on \emph{c}-plane sapphire, according to the procedure described elsewhere\cite{Stefanowicz:2010_PRBa,Bonanni:2011_PRB}. The doped layer is 600~nm thick. The samples are grown under H$_2$ atmosphere, with a pressure of 200~mbar and a temperature of 850$^\circ$C. The precursors used are ammonia (NH$_3$) for nitrogen, trimethylgallium (TMGa) for Ga, dicyclopentadienyl-magnesium (Cp$_2$Mg) for Mg and dicyclopentadienyl-manganese (Cp$_2$Mn) for Mn. The source flow of ammonia was kept constant at 1500~sccm, the TMGa at 5~sccm, and Cp$_2$Mg was varied between 150 and 450 sccm as Cp$_2$Mn was varied between 75 and 490 sccm. The Mn and Mg concentrations considered in this work are both between 0 and 1\% as measured by secondary ion mass spectroscopy (SIMS). 
The absence of parasitic elements like hydrogen or oxygen has been carefully checked with SIMS, energy dispersive x-ray spectropscopy (EDX), Raman spectropscopy, and electron energy loss spectroscopy (EELS). 
Prior to the extensive synchrotron investigations by EXAFS and XES, the structure  of the layers has been characterised by high-resolution x-ray diffraction (HRXRD) on a X'Pert PRO MRD system with a dynamics as high as $10^7$ between the GaN (002) peak and the noise. In addition, high-resolution transmission electron microscopy (HRTEM) was performed on a JEOL 2011 Fast TEM microscope operating at 200 kV and capable of an ultimate point-to-point resolution of 0.19 nm and
allowing to image lattice fringes with a 0.14-nm resolution. The combination of the two techniques has allowed us to rule out the presence of precipitation in the layers.
\newline

	\textbf{EXAFS}: EXAFS spectroscopy has been carried out at the BM08--GILDA Italian beamline\cite{Dacapito:1998_EN} at the ESRF (Grenoble, France). The Mn K edge x-ray absorption spectra have been acquired using a monochromator 
equipped with a pair of Si(311) crystals and run in dynamical focusing mode. 
Harmonics rejection is achieved through a pair of Pd-coated mirrors and the 
monochromator de-tuning. The data are collected in the fluorescence mode using 
a 13-element hyperpure Ge detector and normalized by the incoming flux measured 
with an ion chamber. The incident beam is at 55.7$^\circ$ in respect to the sample surface 
to avoid dichroic effects. The samples are cooled down to liquid nitrogen temperature. 
The counting time and the number of scans for each sample have been chosen in order 
to collect at least 10$^6$ counts per point. The EXAFS signal, $\chi(k)$, is extracted 
from the absorption data, $\mu(E)$, using a smoothing spline algorithm (as implemented 
in the {\sc viper} program) and choosing the energy edge, E$_0$, at the maximum of the
derivative. The data analysis is detailed in the Supplementary Information.
\newline

	\textbf{XES}: Non resonant XES has been measured at the ID26 beamline of the ESRF\cite{Glatzel:2005_CCR}. The optics for the incoming beam consists of three coupled undulators, a double Si crystal monochromator and three Si coated mirrors working at 2.5 mrad incidence for harmonics rejection and beam focusing. The emission spectrometer is run in a vertical Rowland geometry with five Si(110) analyzer crystals working at the (440) reflection, that is, around a Bragg angle of 84.2$^\circ$. The spectrometer energy broadening is approximately 0.9\,eV at Mn\,K-edge (6539\,eV). The experimental geometry consists of the spectrometer and the incoming beam at 90$^\circ$ on the same scattering plane (to minimize the elastic contribution) with the sample surface placed vertically to this plane and at 55.7$^\circ$ incidence angle, that is, the magic angle for wurtzite symmetry in order to avoid dichroism effects due to the linearly polarized beam\cite{Brouder:1990_JPCM}. All the samples are measured at room temperature and are tested against radiation damage.
The emitted fluorescence is scanned around the Mn K$\beta$ main line with the incoming excitation at 6700\,eV.
The quantitative data analysis is based on the integrated absolute values of the difference spectra (IAD) and is performed as a function of the Mg/Mn concentration ratio, $y$. The data have been calibrated by the IAD values obtained from commercial Mn-oxides powders, assuming the ionic approximation and considering the high-spin scenario for both systems. The details of the analysis are reported in the Supplementary Information.
\newline

\textbf{SQUID magnetometry}: The magnetic anisotropy energy density has been assessed by integrating the area between the magnetization curves measured along easy ($H \perp c $) and hard ($H \parallel c$) directions  and expressed as energy per one Mn atom in the given layer. Magnetization curves are measured at 1.85\,K using a superconductor quantum interference device (SQUID), as described previously\cite{Stefanowicz:2010_PRBa,Bonanni:2011_PRB}. The size of the error bars for the anisotropy is determined mostly by the errors related to the inaccuracy of substrate signal compensation and do not include the uncertainty generated by the insufficient strength of the magnetic field to saturate $M$ for hard direction in our SQUID magnetometer (50 kOe).
\newline

\textbf{PL}: 
Photoluminescence is excited with a continuous wave 404\,nm (3.1\,eV) or 442\,nm (2.8\,eV) laser with the excitation power up to tens of mW. The temperature has been varied in the range between 2\,K and 296\,K. An InGaAs type CCD camera coupled to a grating (either 300 grooves/mm or 1200 grooves/mm) monochromator is used as detector. A long wavelength pass filter is placed at the entrance of the monochromator for cutting off the stray laser light. The detection is carried out in the range from 0.7\,eV to 1.5\,eV with a spectral resolution of 0.5\,meV. The integration of the PL signal in Fig.~\ref{fig:PL}b and \ref{fig:PL}d has been performed between 900\,meV and 1100\,meV. The magnetooptical measurements reported in the Supplementary Information are performed in Faraday configuration ($B$\,$\parallel$\,$k$) using a cryostat equipped with a superconducting coil providing a magnetic field up to 10\,T.
\newline

\textbf{Theory -- DFT}:
Calculations for Mn-Mg$_k$ complexes in wz-GaN are performed within the GGA+U approximation using the Quantum Espresso code \cite{Giannozzi:2009_JPCM}. A 96-atoms supercell and  a $3\times3\times3$ Monkhorst-Pack grid for Brillouin zone sampling are employed. The pairing energies are calculated from the following formula:
$ \Delta E$\,$=$\,$E_{\mathrm{tot}}(\mathrm{MnGa}_{47-k}\mathrm{N}_{48}$\,:\,$\mathrm{Mg}_k$\,$+$\,$\mathrm{Ga}_{48}\mathrm{N}_{48}) -E_{\mathrm{tot}}(\mathrm{MnGa}_{47-k+1}\mathrm{N}_{48}:\mathrm{Mg}_{k-1}+\mathrm{Ga}_{47}\mathrm{N}_{48}:\mathrm{Mg}) $
, for $k$ ranging from 1 to 5. From the computed values of the magnetic moment and from the Mn--N distance of every possible complexes, one can obtain, taking into account the relative statistical weight $P_k(y)$ of particular Mn--Mg$_k$ configurations, the average variation of these values as a function of the Mg/Mn ratio. The values of these $P_k(y)$ are then approximated with a binomial law, considering that if Mn can bind up to $m$ Mg atoms ($m$\,$=$\,$3$ in our case), the occurrence probability $P_k(y)$ of particular complexes  Mn-Mg$_k$, at a given ratio of the Mg to Mn concentration $y$, is given by the binomial distribution, $P_k(y) = \binom{m}{k}(y/m)^k(1-y/m)^{m-k}$ for $y \le m$, whereas for $y > m$, $P_m(y) = 1$  and $P_k(y) = 0$ for $k < m$.
The computed values of the local magnetic moment on Mn$_\mathrm{Ga}$ and its nearest neighbouring N atoms as well as its magnitude in the Mn unit cell are collected in the Supplementary Table~S2,  whereas the contour plot of the spin polarization is shown in Fig.~\ref{fig:spin}c.

\section*{Acknowledgements}

The work was supported by the European Research Council through the FunDMS Advanced Grant (\#227690) within the "Ideas" 7th Framework Programme of the EC, by the European Regional Found through grants Innovative Economy Operational Programme 2007-2013 (InTechFun: POIG.01.03.01-00-159/08), by the Austrian Fonds zur {F\"{o}rderung} der wissenschaftlichen Forschung -- FWF (P18942, P20065 and P22477), and by the Polish NCBiR project LIDER. We also acknowledge the European Synchrotron Radiation Facility for provision of synchrotron radiation facilities (proposals HE3609 and 
HS4035), as well as the Interdisciplinary Center of Modeling at the University of Warsaw and the High Performance Computing Center at Texas Southern University for the access to computer facilities.

\section*{Author contributions:}
ThD, under the supervision of AB, initiated the work, fabricated the layers and analyzed the data. MR collected and analyzed XES and XAS spectra under the supervision of PG and FdA, respectively. NGS carried out \emph{ab initio} computations with the assistance of MR, TD, and JAM. Under the supervision of AB, JS and AG carried out optical measurements; TL performed HRTEM; ANQ measured HRXRD; BF did the electrical characterisation; RJ analyzed the samples with SIMS; SD, WS, and MS performed the magnetic studies analyzed by DS and TD. ThD, AB, and TD wrote the manuscript with inputs from all authors. 

\section*{Competing financial interests:} 
The authors declare no competing financial interests.
\end{document}